\documentclass[12pt,preprint]{aastex}
\usepackage{psfig,lscape}
\def\lap{\lower.5ex\hbox{$\; \buildrel < \over \sim \;$}}
\def\gap{\lower.5ex\hbox{$\; \buildrel > \over \sim \;$}}

\def\ergcm2s{${\rm erg\ cm^{-2}\ s^{-1}}$}

\def\ergscm2s{${\rm erg\ cm^{-2}\  s^{-1}}$}

\def\cm-2{${\rm cm^{-2}}$}

\begin{document}

\title{A Soft X-ray Transient in the M31 Bulge}

\author{Benjamin F. Williams\altaffilmark{1}, Michael
R. Garcia\altaffilmark{1}, Jeffrey
E. McClintock\altaffilmark{1}, Frank A. Primini\altaffilmark{1}, and Stephen S. Murray\altaffilmark{1}}
\altaffiltext{1}{Harvard-Smithsonian Center for Astrophysics, 60
Garden Street, Cambridge, MA 02138; williams@head.cfa.harvard.edu;
garcia@head.cfa.harvard.edu; fap@head.cfa.harvard.edu; jem@head.cfa.harvard.edu;
ssm@head.cfa.harvard.edu}


\begin{abstract}

We have examined a probable soft X-ray transient source in the M31
bulge at R.A.=0:42:41.814 $\pm$ 0.08$''$, Dec. = 41:16:35.86 $\pm$
0.07$''$. On the three occasions we observed the source, its spectrum
was soft ($kT_{in} \approx 1$ keV).  The brightest detection of the
source was 2004~July~17 with a 0.3--7 keV luminosity of
$\sim$5$\times10^{37}$ erg s$^{-1}$.  The only previous detection of
the source was in 1979 by the {\it Einstein} observatory.  The
multiple detections over 25 years suggest the duty cycle of the source
is in the range 0.02--0.06.  Coordinated {\it HST} ACS imaging before,
during, and after the outburst revealed no variable optical source
within the position errors of the X-ray source.  The optical data
place a firm upper limit on the brightness of the counterpart of the
X-ray outburst of $B>24.7$, suggesting the binary has a period
$\lap5.2$ days. The X-ray spectrum and lack of bright stars at the
source location indicate the source was a soft transient event
occurring in a low-mass X-ray binary, making this source a good black
hole candidate in M31.

\end{abstract}

\section{Introduction}

Low-mass X-ray binaries (LMXBs) that exhibit bright, soft, transient
X-ray outbursts often contain stellar-mass black hole primaries
(\citealp{mcclintock2004} and references therein).  X-ray observations
of these objects only tell part of the story, as their orbital
periods and masses are typically determined through observations of
their optical counterparts.

Observations with {\it Chandra} and {\it XMM-Newton} have provided
discoveries of many bright transient X-ray sources in M31 in the past
few years
(\citealp{trudolyubov2001,osborne2001,kong2002,williams2004hrc}, and
many others).  However, the optical properties of such events in M31
have received attention only recently.

Combined optical/X-ray studies of X-ray transients constrains their
orbital periods.  Through continued monitoring of the bulge of M31
with the {\it Chandra X-ray Observatory (Chandra)} and {\it Hubble
Space Telescope (HST)}, we have discovered several bright, soft,
transient X-ray sources
\citep{williams2004hrc,williams2005bh1,williams2005bh2,williams2005bh4,williams2005bh6}.
In some cases, the coordinated optical observations of these transient
events has led to the identification of probable optical counterpart
candidates
\citep{williams2004hrc,williams2005bh1,williams2005bh2,williams2005bh6}.
In other cases, the observations have led to meaningful upper limits
on the optical luminosities of the X-ray transient during outburst
\citep{williams2005bh4}.  As the sample of such systems grows, it will
become possible to perform statistical tests, using properties such as
the luminosity and period distribution of the sample to test models of
binary formation and evolution.

Herein we report an outburst from a transient X-ray source in the M31
bulge.  Coordinated optical observations of the source reveal no
variable optical counterpart candidate and show that its optical
counterpart must have been fainter than $B=24.7$.  Section 2 details
our observations and analysis techniques.  Section 3 gives the results
of the analysis.  Section 4 discusses the implications of the results,
and \S~5 provides the conclusions from our work.

\section{Data}

\subsection{X-ray}

We obtained six observations with the {\it Chandra} ACIS-I between
January 2004 and November 2004 relevant to this study.  The
observation identification numbers, dates, pointings, roll angles, and
exposure times of these observations are given in Table~\ref{xobs}.
Although the observations were taken for 5 ks each, the effective
exposure was reduced by $\sim$20\% because the data were taken in
``alternating readout mode'' in order to avoid pileup for any bright
transient source.

All of these observations were reduced in an identical fashion using
CIAO 3.1.  We created exposure maps for the images using the task {\it
merge\_all},\footnote{http://cxc.harvard.edu/ciao/ahelp/merge\_all.html}
and we found and measured positions and count rates of the sources in
the images using the CIAO task {\it
wavdetect}.\footnote{http://cxc.harvard.edu/ciao3.0/download/doc/detect\_html\_manual/Manual.html}.
These data sets each had 3$\sigma$ detection limits of
$\sim$2.2$\times$10$^{-3}$ ct s$^{-1}$, corresponding to 0.3--10 keV
luminosities of $\sim$2$\times$10$^{36}$ erg s$^{-1}$ for a typical
LMXB system in M31.  The background-subtracted count rates are
provided in Table~\ref{xobs}.

We searched our source lists for new sources brighter than
1.5$\times$10$^{37}$ erg s$^{-1}$ that had not appeared in earlier
Chandra images or on recently published lists of X-ray sources in M31.
We focus here on one such source which appeared on 15-Aug-2004 at
R.A.=0:42:41.814, Dec. = 41:16:35.86.  The source is coincident with
an X-ray source seen by the {\it Einstein} HRI in January 1979 called
TF~47 with a 0.2--4 keV luminosity of 1.5$\times$10$^{37}$ erg
s$^{-1}$ \citep{trinchieri1991}.  It has not been detected since that
outburst.  We give this source the formal name CXOM31~J004241.8+411635
and the short name r1-36 using the naming conventions spelled out in
Table~2 of \citet{williams2004hrc}.  This source is 0.65$'$ northwest
of the M31 nucleus.

We aligned the coordinate system of the X-ray images with the optical
images of the Local Group Survey (LGS; \citealp{massey2001}).  The
positions of X-ray sources with known globular cluster counterparts
were aligned with the globular cluster centers in the LGS images using
the IRAF\footnote{IRAF is distributed by the National Optical
Astronomy Observatory, which is operated by the Association of
Universities for Research in Astronomy, Inc., under cooperative
agreement with the National Science Foundation.} task {\it ccmap}.
The precise alignment errors for the observations, which were applied
in determining the X-ray error ellipse, are given in Table~\ref{xpos}.

The position errors of the source on the X-ray images was measured
using the IRAF task {\it imcentroid}.  These errors were all
$<$0.1$''$ and are also provided in Table~\ref{xpos}.  The final
position and errors on the position were calculated by adding the
alignment and position errors for each measurement in quadrature and
then taking the weighted mean of the three independent position
measurements.

Finally, the spectra of r1-36 were extracted using the CIAO task {\it
psextract},\footnote{http://cxc.harvard.edu/ciao/ahelp/psextract.html}
binning the spectrum in energy so that each bin contained $\gap$20
counts.  We then fit each spectrum using CIAO 3.1/Sherpa, trying both
a power-law with absorption and a disk blackbody model with
absorption.  All of the errors reported are 1$\sigma$ unless otherwise
specified.  Errors in the spectral fits were obtained with the {\it
covariance} routine in Sherpa.  Luminosity errors take into account
the normalization errors in the spectral fitting, but count rate
errors do not.  Results are discussed in \S~\ref{results}.

\subsection{Optical}

We obtained three sets of images containing the position of r1-36 with
the Advanced Camera for Surveys (ACS) aboard {\it HST}.  The first
data set, pointed at R.A.=00:42:43.86, Dec.=41:16:30.1 with an
orientation of 55 degrees, was taken on 23-Jan-2004 in an attempt to
detect a different transient source.  This dataset fortuitously
contained the position of r1-36 before it became active.  The second,
pointed at R.A.=00:42:41.50, Dec.=41:17:00.0 with an orientation of
220 degrees, was taken 15-Aug-2004 while the X-ray source was active.
The final observations, pointed at R.A.=00:42:41.50, Dec.=41:17:05.0
with an orientation of 116 degrees, was taken 22-Nov-2004 after the
X-ray source faded.

All three data sets were obtained through the F435W ($B$ equivalent)
filter using the ACS 4-point box dither pattern in order to allow us
to recover the best possible spatial resolution.  The total exposure
time for each set was 2200 s.  Each epoch was combined in an identical
manner using the PyRAF\footnote{PyRAF is a product of the Space
Telescope Science Institute, which is operated by AURA for NASA.} task
{\it multidrizzle},\footnote{multidrizzle is a product of the Space
Telescope Science Institute, which is operated by AURA for
NASA. http://stsdas.stsci.edu/pydrizzle/multidrizzle} which has been
optimized to process ACS imaging data.  The task removes cosmic ray
events and geometric distortions, and it combines the dithered frames
together into one final photometric image in units of counts per
second.  We utilized the capabilities of this task to improve the
sampling of the final images to 0.025$''$ pixel$^{-1}$.

Each final image was aligned with the LGS reference frame using stars
common to both images.  The adjustment of the coordinate systems of
the $HST$ images was performed using the IRAF task {\it ccmap}.  In
all cases, the rms errors of the alignment were smaller than 1 ACS
pixel.  The excellent agreement between the independently aligned
frames can be seen in Figure~\ref{ims}.

Photometry was performed on the relevant sections of the final images
using DAOPHOT~II and ALLSTAR \citep{stetson}.  The count rates
measured (0.25$''$ radius aperture) were converted to VEGA magnitudes
($B$ equivalent) using the conversion formulas provided in the ACS
Data
Handbook\footnote{http://www.stsci.edu/hst/acs/documents/handbooks/DataHandbookv2/ACS\_longdhbcover.html}.

\begin{figure}
\centerline{\psfig{file=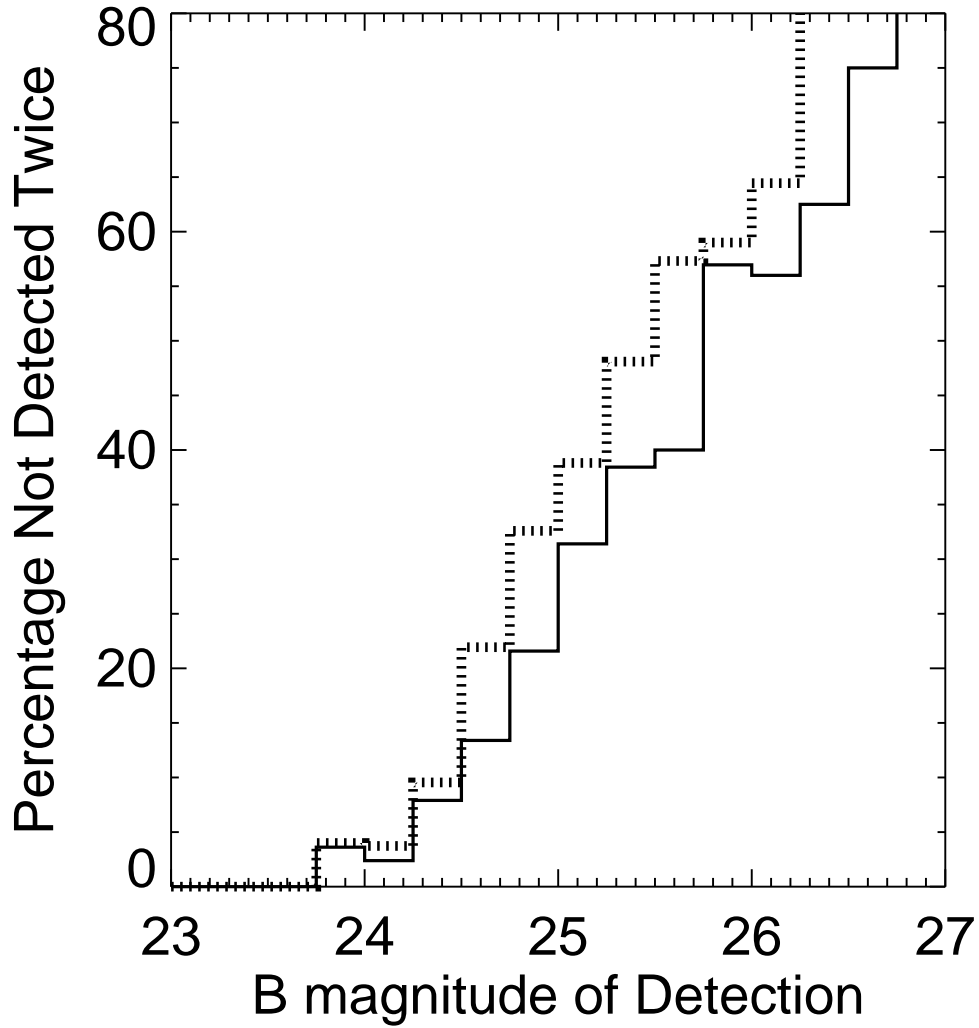,height=6.0in,angle=0}}
\caption{The results of our completeness investigation are shown.
{\it Solid histogram:} Percentage of stars within 9$''$ of r1-36 that
were found by DAOPHOT in the 15-Aug-2004 observation but not in the
22-Nov-2004 observation as a function of $B$-magnitude.  {\it Dashed
histogram:} Percentage of stars within 9$''$ of r1-36 that were found
by DAOPHOT in the 22-Nov-2004 observation but not in the 15-Aug-2004
observation as a function of $B$-magnitude.  The completeness of our
data begins to decrease at $B=24$.}
\label{comp}
\end{figure}

The crowding this close to the M31 nucleus limited the depth to which
we could resolve individual stars.  The limitations of the photometry
were investigated by comparing the results for the same section of sky
in the second and third ACS data sets.  Figure~\ref{comp} shows the
percentage of stars detected in only one data set as a function of $B$
magnitude within 9$''$ of the position of r1-36.  The results show
that confusion begins to limit the completeness of the data at $B=24$,
and the completeness decreases to 50\% at $B\sim25.5$.

\section{Results}\label{results}

\subsection{X-ray}

We determined the position of r1-36 by taking the weighted mean of the
position determination from the three independent detections.  The
errors for each of the three position measurements were calculated by
taking the root-sum-square of the {\it imcentroid} error of the
position on the detector and the {\it ccmap} alignment error with the
LGS coordinate system.  All of these errors are provided in
Table~\ref{xpos}, resulting in a final position and 1$\sigma$ error of
R.A.=0:42:41.814$\pm$0.08$''$, decl.=41:16:35.86$\pm$0.07$''$.  This
position is shown with the 3$\sigma$ error ellipses on
Figure~\ref{ims}.

\begin{figure}
\centerline{\psfig{file=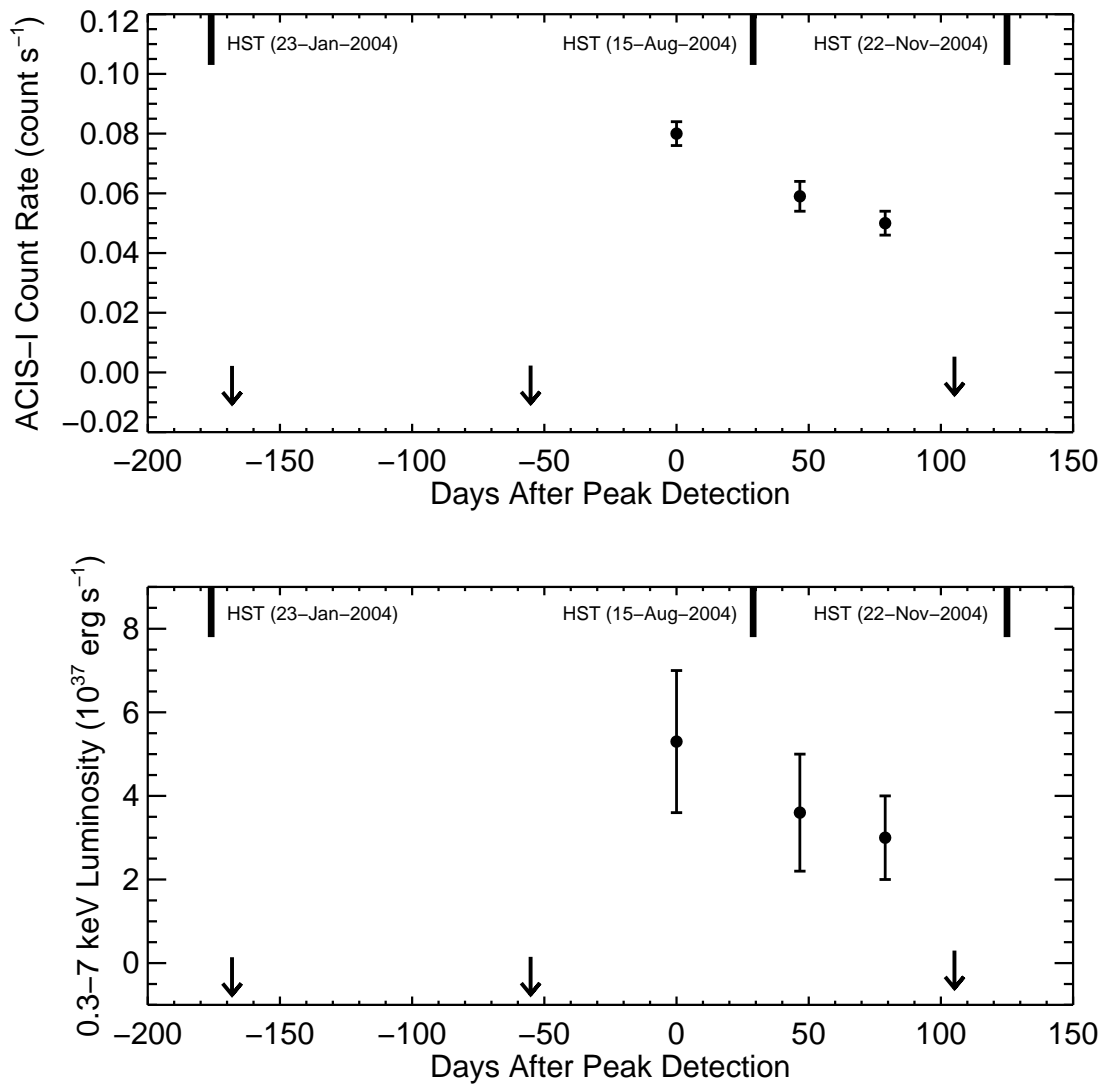,height=6.0in,angle=0}}
\caption{{\it Top:} The X-ray lightcurve for r1-36 is shown.  The
source was strong in X-rays for about 3 months before rapidly fading
to a non-detection.  The times of our coordinated $HST$ observations
are shown with labeled, long vertical tickmarks on the top axis of the
plot. {\it Bottom:} Same as {\it top}, but the y-axis gives the 0.3--7
keV luminosity from disk blackbody spectral fits.  Errors are larger
because they take into account the errors in the spectral fits.}
\label{lc}
\end{figure}

The X-ray lightcurve of r1-36 is shown in Figure~\ref{lc}.  This
lightcurve, which shows a slow decrease in X-ray flux followed by a
sharp drop after 3 months, clearly did not follow a strict exponential
decay (the best fitting exponential has $\chi^2_\nu$=23).  However, in
the 26 days between the final detection and the first non-detection,
the flux decreased by a factor of at least 9.  This final decay
therefore had an $e$-folding decay time of at most 12 days if
exponential decay is assumed.  This type of decay is not unique.
Similar decay patterns have been seen in some Galactic outbursts, such
as the second half of the 1998--1999 outburst of XTE~J1550-564
\citep{sobczak2000}.

Source r1-36 is a second detection of the transient X-ray source
TF~47.  This source has not been detected since the {\it Einstein}
era, when it was detected in only a single epoch of observations with
the HRI (1979~January; \citealp{trinchieri1991}).  Two detected
outbursts of this source over all of the epochs of {\it Einstein},
{\it Rosat}, {\it XMM}, and {\it Chandra} observations constrain the
duty cycle of the transient.  Assuming that the transient was active
for about the same length of time when it appeared in 1979 as it was
during this outburst ($\sim$3 months), the transient was active for at
least 6 months over the past 25 years, giving a lower limit to the
duty cycle of 0.02.  At the same time, all of the {\it Chandra}
observations of M31 to date have observed 1999~October--2002~August
and 2003~November--2004~November without seeing the source active more
than this one time.  The longest gaps in these observations were
$\sim$4 months.  Assuming no outbursts were missed during this
monitoring period, we can infer an upper limit on the duty cycle of
0.06.  Therefore the duty cycle of r1-36 is likely 0.02--0.06.

The X-ray spectra of the three detections of r1-36 were all
well-fitted by absorbed disk blackbody and absorbed power-law models.
They all show that the spectrum is softer than those of high-mass
X-ray binaries, which typically show hard spectra dominated by
power-laws of index closer to 1, The best-fitting indexes were
$2<\Gamma<3$; such values are in the range for LMXBs in the
thermal-dominant state ($\Gamma =$2--5; see Table 4.4 in
\citealp{mcclintock2004}) suggesting that the system is an LMXB
containing an accreting compact object.  The absorption values in
these fits were quite high for soft bright M31 X-ray sources, which
typically have N$_H \sim 1 \times 10^{21}$ cm$^{-2}$
\citep{shirey2001}.  Because the absorption values for the power-law
fits were unusually high, we did not consider a power-law the
appropriate model for fitting our data from r1-36.  However, for
completeness, the results from the power-law fits are given in
Table~\ref{pl}.

In the disk blackbody fits (see Table~\ref{spectab1}), the measured
absorption was within the errors of the known foreground absorption
toward M31 [5$\times$10$^{20}$ cm$^{-2}$ for $A_V=0.3$
\citep{predehl1995,schlegel1998}].  We therefore fixed the absorption
to this value for all three detections.  These fits yielded
temperatures similar to those of Galactic LMXBs in outburst
\citep{mcclintock2004} and indicated a decrease in temperature as the
X-ray flux faded.  The best-fitting parameter values, along with the
improved fitting statistics, suggest that a disk blackbody is the
physical process responsible for most of the X-ray emission from
r1-36.

We provide all of the details of the disk blackbody spectral fits in
Table~\ref{spectab1}.  The fits yield an unabsorbed 0.3--7 keV X-ray
luminosity during the 15-Aug-2004 {\it HST} observation of
$\sim$5$\times10^{37}$ erg s$^{-1}$.  This luminosity does not shed
light as to the nature of the accreting compact object because it is
below the Eddington limit for both neutron stars and black holes.
However, the observed decay of the X-ray flux and disk temperature are
hallmarks of the decline of black hole transients in the thermal
dominant state
(e.g. \citealp{ebisawa1994,sobczak2000,park2004,mcclintock2004}).

\subsection{Optical}

\begin{figure}
\centerline{\psfig{file=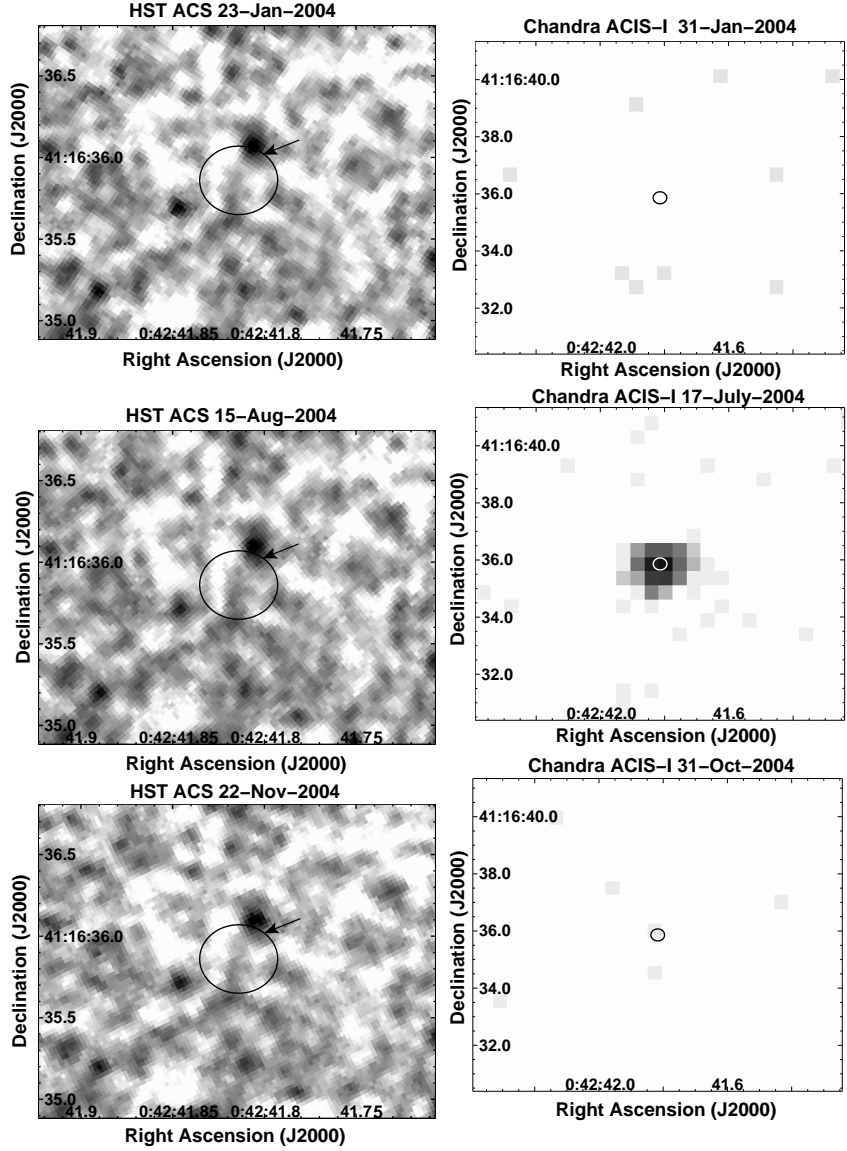,height=6.0in,angle=0}}
\caption{{\it Left panels: HST} ACS images of the location of r1-36
taken before, during, and after the outburst.  The black ellipses mark
the 3$\sigma$ error for the X-ray position of r1-36.  The small black
arrows mark the brightest star in the error circle that was detected
during the 15-Aug-2004 observation. {\it Right panels: Chandra} ACIS-I
images of r1-36 before, during, and after the outburst.  The ellipses
in the left panels correspond to the ellipses on the X-ray images.}
\label{ims}
\end{figure}

The coordinated $HST$ images of the region of interest are shown on
the left side of Figure~\ref{ims}.  Each image is shown next to the
most contemporaneous X-ray image, and the 3$\sigma$ position error
ellipse for the position of r1-36 is drawn on each image.  The
analysis of these images is severely hampered by crowding, as
demonstrated by the unresolved structure in all three images.  There
is no obvious bright star in the error ellipse, consistent with the
X-ray spectrum in suggesting that the source is an LMXB.

The DAOPHOT analysis of the optical data did not reveal any resolved
stars inside the error ellipse brighter than $B=24.7$.  No star in or
near the error ellipse shows a significant increase in brightness
during the epoch when the X-ray source was active.  The only star
detected in the 15-Aug-2004 observation (when r1-36 was active) inside
the error ellipse is marked with the arrows on Figure~\ref{ims} at
R.A.=00:42:41.803, Dec.=41:16:36.00. This sole resolved star is
blended with the bright $B=23.6$ star 0.1$''$ to the north, outside
the error ellipse.  The star inside the ellipse was measured to be
brightest in the 15-Aug-2004 observation, with $B=24.70\pm0.08$.  This
star was also detected in the 23-Jan-2004 observation (when r1-36 was
quiescent), but it was slightly fainter ($B=24.92\pm0.11$).  It was
not detected by DAOPHOT in the 22-Nov-2004 observation.

The low-significance brightness increase and disappearance of this
star make it the only optical counterpart candidate for r1-36, but the
variability detection is not reliable.  The increase in brightness is
only 1.6$\sigma$.  Furthermore, the results of our completeness study,
shown in Figure~\ref{comp}, indicate that the completeness begins to
decrease at $B\sim24$, and is only $\sim$80\% at $B=24.7$, providing a
1/5 chance that such a star will be missed in the analysis.  Therefore
the fact that it was missed in 1 out of 3 images is not unexpected.
On the other hand, as this was the only star detected in the 3$\sigma$
error ellipse in the 15-Aug-2004 data, it provides a reliable
upper-limit of $B\geq24.7$ for the optical brightness of the X-ray
transient during outburst.

For the purpose of determining an upper limit to the optical
luminosity of the outburst, we assume that the counterpart candidate
with $B=24.70\pm0.08$ was the counterpart.  With N$_H =
5\times10^{20}$ cm$^{-2}$ (see Table~\ref{spectab1}), applying the
relation of \citet{predehl1995} and a standard extinction law, we find
$A_B = 0.4$.  Assuming $m$-M=24.47, the absolute $B$ magnitude was
$B=-0.17\pm0.08$.  Assuming an intrinsic $B-V$ color of $-0.09\pm0.14$
[the mean of the Galactic LMXB catalog of \citet{liu2001}], we find
$M_V=-0.08\pm0.16$, or an upper-limit of $M_V\geq-0.24$.

Our measurements allow us to provide a rough prediction of the orbital
period of r1-36 using the empirical relation between X-ray luminosity,
optical luminosity, and orbital period for Galactic LMXBs determined
by \citet{vanparadijs1994}.  This relation has been tested for more
recent transient events and for observations separated by up to 3
weeks by \citet{williams2005bh1,williams2005bh4}, showing that it
provides reliable orbital period predictions even for events with
complex multiwavelength lightcurves.  In addition, since r1-36 shows a
decay curve reminiscent of the second-half of the 1998-1999 outburst
of XTE~J1550-564, we tested the relation for this case.  The second
half of that outburst had an X-ray luminosity of $\sim2\times 10^{38}$
erg s$^{-1}$ (for a distance of 5.3$\pm$2.3 kpc;
\citealp{orosz2002j1550}), and the optical counterpart showed
$V\sim16.3$.  Applying the extinction of $A_V=4.75$
\citep{orosz2002j1550} gives M$_V = -2.1^{+1.2}_{-0.8}$.  These
numbers would yield a period prediction of $\gap$1 day.  This limit is
correct, as the true period is 1.55 days.

In the case of r1-36, if we insert the optical luminosity of
$M_V=-0.08\pm0.16$ and the unabsorbed 0.3--7 keV X-ray luminosity of
5$\times$10$^{37}$ erg s$^{-1}$ into the \citet{vanparadijs1994}
relation, we obtain a prediction for the orbital period of
1.7$^{+3.5}_{-1.0}$ days. Therefore, assuming r1-36 is an LMXB similar
to those in our own Galaxy and that the true counterpart was no
brighter than the only star detected within the 3$\sigma$ error
ellipse during the outburst, the predicted upper-limit for the period
of the system is $\lap$5.2 days.

\section{Discussion: Source Classification}

Ideally we would like to fully classify the transient X-ray source
r1-36.  While its observed properties show that it is an LMXB in M31,
the current observations do not allow a final conclusion to be drawn
as to whether the primary member of the binary is a neutron star or
black hole.  Nevertheless, the similarities between r1-36 and X-ray
binaries known to contain black holes make it a good black hole
candidate.

First of all, the source is most likely in M31.  At the flux levels
observed ($\sim$5$\times$10$^{-13}$ erg cm$^{-2}$ s$^{-1}$), the
number of background sources is $<$1 per square degree (extrapolating
the results of \citealp{hasinger1998} to brighter fluxes).  With our
ACIS field covering $\sim$0.07 deg$^2$, the probability that this
source is in the background is $<$7\%.  The source is also unlikely to
be in the foreground.  The disk blackbody spectral fits provide a
normalization value that, given the distance to the source, indicates
$R_{in}cos^{1/2}(i)$ (where $R_{in}$ is the inner radius of the
accretion disk in km and $i$ is the inclination of the disk).  If the
source is at the distance of M31 (780 kpc), $R_{in} \sim
10/cos^{1/2}(i)$ km, physically reasonable for a compact object.  On
the other hand, if the source is in the foreground (10 kpc), $R_{in}
\sim 0.1/cos^{1/2}(i)$ km, quite small for any accreting object, even
for high inclination angles.  Together, these results suggest the
source is in M31.

The source is likely an LMXB, as suggested by the significant
variability, spectrum, and lack of high-mass stellar counterpart
candidates.  The current data set allows either a neutron star or
black hole primary.  There are spectral characteristics that could
discern between the two, such as a blackbody component indicative of a
neutron star envelope \citep{tanaka1996}.  Unfortunately, our spectra
are not of sufficient quality to reliably detect such additional
components.  Our spectra of r1-36 are soft, are well-fitted by a disk
blackbody model, and indicate a peak luminosity of
$\sim$5$\times$10$^{37}$ erg s$^{-1}$.  LMXBs with black hole
primaries and those with neutron star primaries can both have soft
disk blackbody spectral components and peak luminosities of
$\sim$10$^{37}$--10$^{38}$ erg s$^{-1}$ \citep{tanaka1996}.  Our X-ray
spectra therefore allow the presence of either type of compact object.

While we cannot be certain about the nature of the compact object in
r1-36, it certainly could contain a black hole.  Most known transient
X-ray binaries containing neutron stars are HMXBs with Be secondaries.
Source r1-36 has a softer spectrum and a fainter secondary than such
HMXB transients.  In addition, more than half of Galactic transient
LMXBs contain black holes
\citep{mcclintock2004,tanaka1996}. Furthermore, with the exception of
GRS 1915+105, which has remained active since its discovery in 1992,
all LMXBs with dynamically-confirmed black hole primaries are
transient sources \citep{mcclintock2004}.

Finally, the X-ray decay pattern of r1-36 mimics typical decays of
black hole binaries but differs from typical decays of neutron star
binaries.  The inner disk temperature of r1-36 decays from 1.2 keV to
0.8 keV; similarly, black hole binaries in the thermal dominant state
have very comparable disk temperatures (e.g., see Table 4 in
\citealp{mcclintock2004}) that invariably decay smoothly and
monotonically (e.g.,
\citealp{ebisawa1994,sobczak1999,sobczak2000,park2004}).  In contrast,
neutron star binaries appear to exhibit constant or non-monotonic
temperatures as they decay in the soft state.  For example, Aquila X-1
showed a constant inner disk temperature of $\sim$2.3 keV during the
thermal dominant phase of its 2000 outburst \citep{maitra2004}, and
the temperature of 4U 1630-47 first declined rapidly and thereafter
fluctuated irregularly around a value of $\sim$0.7 keV during most of
its decline phase \citep{tomsick2000}.  These comparisons indicate
that r1-36 has more in common with known black hole binaries than with
neutron star binaries.

Succinctly, our data set cannot reliably determine the nature of the
primary in the M31 LMXB X-ray transient r1-36.  On the other hand, the
data show that r1-36 is not similar to most X-ray binaries known to
contain neutron stars, which are persistent and/or have high-mass
secondaries.  The data also show that r1-36 {\it is} similar to most
X-ray binaries known to contain black holes, which are transient, have
low-mass secondaries, and exhibit similar luminosity and temperature
decay patterns.  These similarities make r1-36 a good black hole
candidate in M31.

\section{Conclusions}

We have studied a recent outburst of the X-ray transient source TF~47
\citep{trinchieri1991}, which we have detected on three occasions with
{\it Chandra}.  We call this most recent outburst
CXOM31~J004241.8+411635, or r1-36.  This second detected outburst from
this source suggests its duty cycle is in the range 0.02--0.06.

The soft X-ray spectrum of the source indicates that it is likely an
LMXB.  Spectral fits suggest the X-rays came from a hot accretion disk
with an inner edge temperature of $\sim$1 keV.  The
absorption-corrected 0.3--7 keV luminosity of the source during the
outburst was $\sim$5$\times$10$^{37}$ erg s$^{-1}$. 

$HST$ ACS observations before, during, and after the outburst set a
limit of $B<24.7$ for the counterpart to the system.  This fact
supports the argument that the secondary is low-mass.  The X-ray
luminosity and optical upper-limit during outburst suggest the period
of the binary system is $\lap$5.2 days.

The X-ray spectrum and optical data show that r1-36 is a transient
LMXB in M31. This evidence does not allow a reliable conclusion as to
the nature of the primary, as neither the spectrum nor the luminosity
of r1-36 rule out the possibility that the accreting compact object is
a neutron star.  In the Galaxy, these types of sources have a higher
probability of containing a black hole than do HMXBs or persistent
bright X-ray sources.  In addition, the X-ray decay properties of
r1-36 bear a stronger resemblance to black hole binaries than to
neutron star binaries. Therefore, while there is no conclusive
evidence that r1-36 contains a black hole, it is a good black hole
candidate.

Out of 6 transient X-ray sources likely to be LMXBs whose optical
luminosities have been studied so far
\citep{williams2004hrc,williams2005bh1,williams2005bh2,williams2005bh4,williams2005bh6},
this is the fourth one with a predicted orbital period estimate of
less than a few days.  The three others are s1-86
\citep{williams2005bh1}, r2-71 \citep{williams2005bh4}, and r3-127
\citep{williams2005bh6}.  Together, these results begin to hint that
at least half of the transient LMXB sources in the M31 bulge may be
rather short period systems with small accretion disks.  While the
statistics of a sample of 6 are clearly limited, the orbital period
distribution (inferred from the X-ray/optical luminosity ratio
distribution) does not appear grossly different from that of similar
systems in the Galaxy (e.g. \citealp{orosz2002}).

Support for this work was provided by NASA through grant number
GO-9087 from the Space Telescope Science Institute and through grant
number GO-3103X from the {\it Chandra} X-Ray Center.  MRG acknowledges
support from NASA LTSA grant NAG5-10889.  SSM acknowledges the support
of the HRC contract NAS8-03060.  JEM acknowledges the support of NASA
grant NNG0-5GB31G.

\begin{deluxetable}{ccccccccccc}
\tablecaption{{\it Chandra} ACIS-I observations}
\tableheadfrac{0.01}
\tablehead{
\colhead{{\footnotesize ObsID}} &
\colhead{{\footnotesize Date}} &
\colhead{{\footnotesize R.A. (J2000)}} &
\colhead{{\footnotesize Dec. (J2000)}} &
\colhead{{\footnotesize Roll (deg.)}} &
\colhead{{\footnotesize Exp.\tablenotemark{a}}} &
\colhead{{\footnotesize Cts}\tablenotemark{b}} &
\colhead{{\footnotesize Rate}\tablenotemark{c}}
}
\tablenotetext{a}{The exposure time of the observation in ks.}
\tablenotetext{b}{The number of background-subtracted counts from r1-36 in the detection.  Non-detections are given a value of 9 counts for a 3$\sigma$ upper limit.}
\tablenotetext{c}{The count rate of r1-36 in units of ct s$^{-1}$.}
\startdata
4681 & 31-Jan-2004 & 00 42 44.4 & 41 16 08.3 & 305.55 & 4.1 & $<$9 & $<$0.0022\\
4682 & 23-May-2004 & 00 42 44.4 & 41 16 08.3 & 79.99 & 3.9 & $<$9 & $<$0.0023\\
4719 & 17-Jul-2004 & 00 42 44.3 & 41 16 08.4 & 116.83 & 4.1 & 328 & 0.080$\pm$0.004\\
4720 & 02-Sep-2004 & 00 42 44.3 & 41 16 08.4 & 144.80 & 4.1 & 161 & 0.059$\pm$0.005\\
4721 & 04-Oct-2004 & 00 42 44.3 & 41 16 08.4 & 180.55 & 4.1 & 207 & 0.050$\pm$0.004\\
4722 & 31-Oct-2004 & 00 42 44.3 & 41 16 08.4 & 225.99 & 3.9 & $<$9 & $<$0.0053\\
\enddata
\label{xobs}
\end{deluxetable}

\begin{deluxetable}{cccccccccc}
\tablecaption{Position Measurements and Errors of r1-36}
\tableheadfrac{0.01}
\tablehead{
\colhead{{ObsID}} &
\colhead{{R.A. (J2000)}} &
\colhead{{$\sigma_{pos}$\tablenotemark{a} ($''$)}} &
\colhead{{$\sigma_{AL}$\tablenotemark{b} ($''$)}} &
\colhead{{$\sigma_{tot}\tablenotemark{c}$ ($''$)}} &
\colhead{{Dec. (J2000)}} &
\colhead{{$\sigma_{pos}$ ($''$)}} &
\colhead{{$\sigma_{AL}$ ($''$)}} &
\colhead{{$\sigma_{tot}$ ($''$)}}
}
\tablenotetext{a}{Random position errors were measured using the IRAF task {\it
imcentroid}.}
\tablenotetext{b}{Errors in the alignment between the X-ray image and the LGS coordinate system were measured using the IRAF task {\it ccmap}.}
\tablenotetext{c}{Total position errors were calculated by adding the position and alignment errors in quadrature.}
\tablenotetext{d}{The mean position and errors were calculated using standard statistics for combining multiple measurements \citep{bevington}.}
\startdata
4719 & 0:42:41.813 & 0.06 & 0.12 &  0.13 & 41:16:35.77 & 0.06 & 0.12 &  0.13\\
4720 & 0:42:41.816 & 0.08 & 0.14 &  0.16 & 41:16:35.82 & 0.08 & 0.07 &  0.11\\
4721 & 0:42:41.812 & 0.07 & 0.14 &  0.16 & 41:16:36.01 & 0.07 & 0.10 &  0.13\\ 
\hline
Mean\tablenotemark{d} & 0:42:41.814 & & & 0.08 & 41:16:35.86 & & & 0.07\\
\enddata
\label{xpos}
\end{deluxetable}

\clearpage

\begin{deluxetable}{ccccccccc}
\tablecaption{Power-law Spectral Fits to r1-36}
\tableheadfrac{0.01}
\tablehead{
\colhead{{ObsID}} &
\colhead{{Range\tablenotemark{a} (keV)}} &
\colhead{{$N_H$\tablenotemark{b}}} &
\colhead{{$\Gamma$\tablenotemark{c}}} &
\colhead{{$\chi^2/dof$\tablenotemark{d}}} &
\colhead{{$Q$\tablenotemark{e}}} &
\colhead{{$L_X$\tablenotemark{f}}}
}
\tablenotetext{a}{The energy range over which the spectrum was fitted.}
\tablenotetext{b}{The absorption column, in units of 10$^{20}$ cm$^{-2}$.  Values with no errors are from fits with fixed absorption.}
\tablenotetext{c}{The best-fitting photon index of the spectrum.}
\tablenotetext{d}{The quality of the fit is given by showing the value of $\chi^2$ and the number of degrees of freedom ($dof$) in the fit.}
\tablenotetext{e}{The probability that the fitted model represents the true intrinsic spectrum (based on the value of $\chi^2/dof$).}
\tablenotetext{f}{The absorption-corrected 0.3--7 keV luminosity of the source in units of $10^{37}$ erg s$^{-1}$, taking into account the uncertainties in the spectral fit.}
\startdata
4719 & 0.35--5 & 19$\pm$8 & 2.0$\pm$0.2 & 8.93/12 & 0.71 & 7.7$\pm$1.5\\
4720 & 0.35--3.5 & 36$\pm$13 & 2.5$\pm$0.3 & 6.43/5 & 0.27 & 8.0$\pm$2.4\\
4721 & 0.35--3.5 & 39$\pm$11 & 3.0$\pm$0.3 & 5.44/7 & 0.61 & 2.5$\pm$0.7\\
\enddata
\label{pl}
\end{deluxetable}

\begin{deluxetable}{ccccccccccc}
\tablecaption{Disk Blackbody Spectral Fits to r1-36 with N$_H = 5 \times 10^{20}$ cm$^{-2}$}
\tableheadfrac{0.01}
\tablehead{
\colhead{{ObsID}} &
\colhead{{\footnotesize{Range (keV)\tablenotemark{a}}}} &
\colhead{{$N_H$\tablenotemark{b}}} &
\colhead{{T$_{in}$\tablenotemark{c} (keV)}} &
\colhead{{\footnotesize R$_{in}$ cos$^{1/2}(\theta)$\tablenotemark{d}}} &
\colhead{{$\chi^2/dof$}\tablenotemark{e}} &
\colhead{{$Q$}\tablenotemark{f}} &
\colhead{{\footnotesize $L_X$\tablenotemark{g}}}
}
\tablenotetext{a}{The energy range over which the spectrum was fitted.}
\tablenotetext{b}{The absorption column, in units of 10$^{20}$ cm$^{-2}$.  Values with no errors are from fits with fixed absorption.}
\tablenotetext{c}{The temperature of the inner edge of the accretion disk.}
\tablenotetext{d}{The radius of the inner edge of the accretion disk in km multiplied by the square-root of the cosine of the inclination of the disk, assuming the source is at a distance of 780 kpc.}
\tablenotetext{e}{The quality of the fit is given by showing the value of $\chi^2$ and the number of degrees of freedom ($dof$) in the fit.}
\tablenotetext{f}{The probability that the fitted model represents the true intrinsic spectrum (based on the value of $\chi^2/dof$). }
\tablenotetext{g}{The absorption-corrected 0.3--7 keV luminosity of the source in units of $10^{37}$ erg s$^{-1}$, taking into account the uncertainties in the spectral fit.}
\startdata
4719 & 0.35--5 & 0$\pm$5 & 1.2$\pm$0.1 & 10$\pm$3 & 7.24/12 & 0.84 & 5.1$\pm$1.9\\
4719 & 0.35--5 & 5 & 1.2$\pm$0.1 & 10$\pm$3 & 8.09/13 & 0.84 & 5.3$\pm$1.7\\
4720 & 0.35--3.5 & 7$\pm$8 & 1.0$\pm$0.1 & 13$\pm$3 & 5.47/5 & 0.36 & 3.7$\pm$2.1\\
4720 & 0.35--3.5 & 5 & 1.0$\pm$0.1 & 12$\pm$3 & 5.56/6 & 0.47 & 3.6$\pm$1.4\\
4721 & 0.35--3.5 & 5$\pm$6 & 0.79$\pm$0.09 & 18$\pm$5 & 4.02/7 & 0.78 & 3.0$\pm$1.5\\ 
4721 & 0.35--3.5 & 5 & 0.80$\pm$0.07 & 18$\pm$4 & 4.02/8 & 0.85 & 3.0$\pm$1.0\\
\enddata
\label{spectab1}
\end{deluxetable}

\clearpage

\end{document}